\documentclass[12pt]{iopart}

\usepackage{iopams}
\usepackage{graphicx}
\usepackage{dcolumn}
\usepackage{bm}
\usepackage{color}
\usepackage{theorem}

\newtheorem{thm}{Theorem}

\newcommand{\hubble}{\ensuremath{\mathrm{H}}}
\newcommand{\deviation}{\scalebox{0.7}[0.8]{\ensuremath{\Sigma}}}
\newcommand{\pressure}{\ensuremath{\mathrm{p}}}
\newcommand{\energy}{\ensuremath{\mathrm{\rho}}}
\newcommand{\state}{\ensuremath{\mathrm{w}}}
\newcommand{\isoredshift}{\ensuremath{\mathsf{z}}}
\newcommand{\factor}{\ensuremath{\mathsf{R}}}
\newcommand{\factorDelta}{\ensuremath{\Delta \mathcal{R}}}

\begin{document}
\title{The nonlinear patterns of the cosmic anisotropy in the late time universe.}

\author{Leandro G. Gomes}
\address{Federal University of Itajub\'a, Av. BPS, 1303, Itajub\'a-MG, 37500-903, Brazil}
\eads{\mailto{lggomes@unifei.edu.br}}

\date{\today}

\begin{abstract}
In this manuscript, we investigate the patterns satisfied by the cosmological anisotropy under the hypothesis of the observers being co-moving with a perfect fluid whose induced space sections are homogeneous with vanishing scalar curvature. We describe the positive increment $\factorDelta$ that the Hubble parameter in the anisotropic model will have as it is compared to its isotropic counterpart. In general, it has an exponential awakening at a specific redshift $\isoredshift_A$. 
We also show that the deceleration and the jerk along the principal directions of the anisotropy are constrained by simple algebraic equations that do not depend on the type of matter present. These characteristic patterns form a valuable framework to distinguish the cosmological anisotropy from any other kind, thus adding an useful tool to probe its upper limits in the supernovae surveys, which are orders of magnitude away from those observed in the CMB.    
\end{abstract}

\maketitle




%
%
%
%
%

In a general spacetime $M$, let us distinguish the class of "cosmological" observers as the flow lines of a time-like, vorticity-free fundamental $4$-velocity $u$. Let $\Theta_\mu^\nu$ denote the associated expansion tensor, such that their eigeinvectors form a space-like orthonormal frame $\{e_1,e_2,e_3\}$ \cite{ellis_mac_marteens}. At each point, their eingeinvalues are the Hubble parameters, $\hubble_1$, $\hubble_2$ and $\hubble_3$, and $\hubble(\hat{n})=(n^1)^2\, \hubble_1+(n^2)^2\, \hubble_2+(n^3)^2\, \hubble_3$ is the Hubble factor seen at the direction $\hat{n} = n^i\, e_i$, $||\hat{n}||=1$. These principal directions are, in principle, distinguishable: one axis stands for the maximum of $\hubble(\hat{n})$, the other for its minimum and the third is uniquely chosen by orthogonality and orientation-preserving. It is convenient to represent this scheme by the isotropic Hubble parameter, $\hubble = \Theta_{\mu}^{\mu}/3 = \left( \hubble_1 +  \hubble_2 + \hubble_3 \right)/3$, the anisotropy magnitude $\deviation$ ("shear parameter"\cite{wainwright}, "Cosmic Shear"\cite{Tedesco}) and  the anisotropy polar angle $\alpha$:
\begin{equation}\label{Eq:DeterminationDeviationPhase}
\deviation = 
\sqrt{\frac{1}{6}\, \sum_{k=1}^3\,\left(\frac{\hubble_k-\hubble}{\hubble}\right)^2} 
\quad \textrm{and} \quad
\sin \alpha=\frac{\hubble_1-\hubble}{2\, \deviation \, \hubble} \, .
\end{equation}
Following just like in the reference \cite{BGK}, where $\sigma_k=\hubble_k-\hubble$, we have\footnote{Some authors attribute a signal to $\deviation$. Here we put it in the signal of $\sin\alpha$. }
\begin{equation}\label{Eq:AnisotropicHubble}
\hubble_k = \left(1 + 2\, \deviation \, \sin\alpha_k  \,\right)\,\hubble 
\qquad \alpha_k = \alpha + \frac{2(k-1)\pi}{3}\, .
\end{equation}
The generalized Friedmann equation is \cite{ellis_mac_marteens}
\begin{equation}\label{Eq:Friedmann}
3\,\left(1 - \deviation^2 \right)\, \hubble^2 =  \energy - \overline{R}/2  \, ,
\end{equation}
where $\overline{R}$ is the scalar curvature of the corresponding space section orthogonal to $u$ and $\energy=T_{\mu\nu}u^\mu u^\nu$ the energy density of the matter component seen by the $u$-observers. As we assume the positivity of $\energy - \overline{R}/2$, $\deviation$ measures the amount of anisotropy in a scale that varies from the isotropic point, $\deviation =0$, to the maximally anisotropic point, $\deviation = 1$:
\begin{equation}\label{Eq:IsotropicScale}
\textrm{(Isotropy)} \quad 0 \quad \leftarrow 
\quad \deviation \quad  
\rightarrow \quad 1  \quad \textrm{(Maximum anisotropy).} 
\end{equation}

The deceleration and the jerk appear in the expansion of the luminous distance against the redshift, just like as if we were considering each past-directed null geodesic arriving at the event of observation $p_0 \in M$ as its counterpart in the standard model of Cosmology\cite{Weinberg}. Hence, fixed the spatial direction $\hat{n}$ at $p_0$, we have
\begin{equation}\label{Eq:LuminousDistance}
 d_L(z,\hat{n})= \frac{1}{\hubble_0(\hat{n})}\left[\, z + \frac{1-q_0(\hat{n})}{2!}\, z^2 
 +  \frac{A_3(\hat{n})-j_0(\hat{n})}{3!}\, z^3 + O(z^4)\,\right] \, ,
\end{equation}
where $A_3(\hat{n})=q_0(\hat{n})+ 3 \, q_0(\hat{n})^2 - 1- \overline{R}_0/(6\, \hubble_0(\hat{n})^2)$. For spaces of constant sectional curvature $K=-1,0,1$, $\overline{R}(p_0)/6$ reduces to $K/a_0^2$. Needless to say, each cosmographic parameter considered so far in this manuscript is not only model dependent, but they also vary with the spatial location of the observer in the inhomogeneous models. Hence, in those cases, there is the necessity of distinguishing the "cosmological" observers from the others \cite{BGS}, as for example, those who perceive an almost isotropic background radiation.

Let us assume the observers to be co-moving with a perfect fluid, $T_{\mu\nu}= \energy\, u_\mu u_\nu+ \pressure g_{\mu\nu}$. In order to consider the effects of the anisotropy only, we will suppress any inhomogeneity in our model, and to be consistent with the observations \cite{Linder}, we take the simplest choice for the curvature, $\overline{R}=0$. These conditions are possible only in a diagonal Bianchi-I model, as any other spatially homogeneous possibility, like the flat RW\cite{ellis_mac_marteens} or the LRS Bianchi-I\cite{wainwright} or the non-tilted LRS Bianchi-$VII_0$\cite{Ellis_MacCallum,Tsoubelis}, can be seen as a special case of the first. Hence, taking advantage of this simple scenario, we can follow close to the presentation of the Einstein's equations as in \cite{BGK}, thus integrating $d\deviation/d\energy=\dot{\deviation}/\dot{\energy}$,  where $\dot{\deviation}=-\,3\, \deviation \, \hubble\, \left(\energy-\pressure \right)(1-\deviation^2)/(2\energy)$ and $\dot{\energy}=-3\, \hubble\, (\energy+\pressure)$, which gives us
\begin{equation}\label{Eq:DeviationEnergyRepresentation}
\deviation (\energy)=  \frac{ \deviation_0}{\sqrt{\deviation_0^2 + (1-\deviation_0^2)\, e^{-\, G(\energy)}\,}} 
\qquad 
G(\energy)= \int_{\energy_0}^{\energy} \, \frac{\energy'-\pressure (\energy')}{\energy'+\pressure(\energy')} \, \frac{d\energy'}{\energy'} \, .
\end{equation}
As we define the  isotropic redshift by $\isoredshift+1:=a_0/a$, $a$ the isotropic scale factor obtained from $\hubble=\dot{a}/a$, the conservation equation turns out to be
\begin{equation}\label{Eq:conservationZ}
 \frac{d\isoredshift}{\isoredshift+1}=\frac{d\energy}{3\, (\energy+\pressure)} \, .
\end{equation}
%
%
%
Hence, using $\isoredshift$ as a time variable, we can define
\begin{equation}\label{Eq:FactorR}
1+ \factor(\isoredshift,\deviation_0) 
 =  \frac{\hubble (\isoredshift)}{\hubble_{0,I}}
 = \sqrt{\frac{\energy(\isoredshift)}{(1-\deviation(\isoredshift)^2)\, \energy_0}} \, ,
\end{equation}
where $\hubble_{0,I}=\hubble_{0}\, \sqrt{1-\deviation_0^2}=\sqrt{\energy_0/3}$ is the Hubble constant seen from the isotropic point of view. As we note that $\factor(\isoredshift,\deviation_0)>\factor(\isoredshift,0)$ for $\deviation_0 \ne 0$, we arrive to our first constraint on the cosmological anisotropy:
\begin{thm}\label{Thm:ConstraintHubble}
On the hypothesis of the observers being co-moving with a perfect fluid along with the scalar-flat and homogeneous space sections, if $\deviation_0 \ne 0$, then the anisotropy in the Hubble parameter will define a positive increment $\factorDelta\,\, \hubble_{0,I}$ in $\hubble(\isoredshift)$ as it is compared to its isotropic counterpart, where     
\begin{equation}\label{Eq:HubbleIncrement}
\factorDelta(\isoredshift,\deviation_0) 
 = \factor(\isoredshift,\deviation_0)-\factor(\isoredshift,0)
\end{equation}
and $\factor(\isoredshift,\deviation_0)$ is determined by the equations (\ref{Eq:DeviationEnergyRepresentation}), (\ref{Eq:conservationZ}) and (\ref{Eq:FactorR}).
\end{thm}

For a better appreciation of the theorem \ref{Thm:ConstraintHubble}, let us take the linear equation of state $\pressure=\state_0\, \energy$, $\state_0$ a constant. In this case, we have
\begin{equation}\label{Eq:DeviationRedshiftRepresentation}
\deviation (\isoredshift)=  \frac{ \deviation_0\, (1+\isoredshift)^{\frac{3}{2}\,(1-\state_0)}}{\sqrt{1+ ((1+\isoredshift)^{3\,(1-\state_0)}-1)\, \deviation_0^2}} \, .
\end{equation}
\begin{table}
 	\label{Tab:AnisotropyVsRedshift}
\begin{center}
    \begin{tabular}{cc}
    	    \includegraphics[width=0.45\textwidth,height=0.4\textwidth]{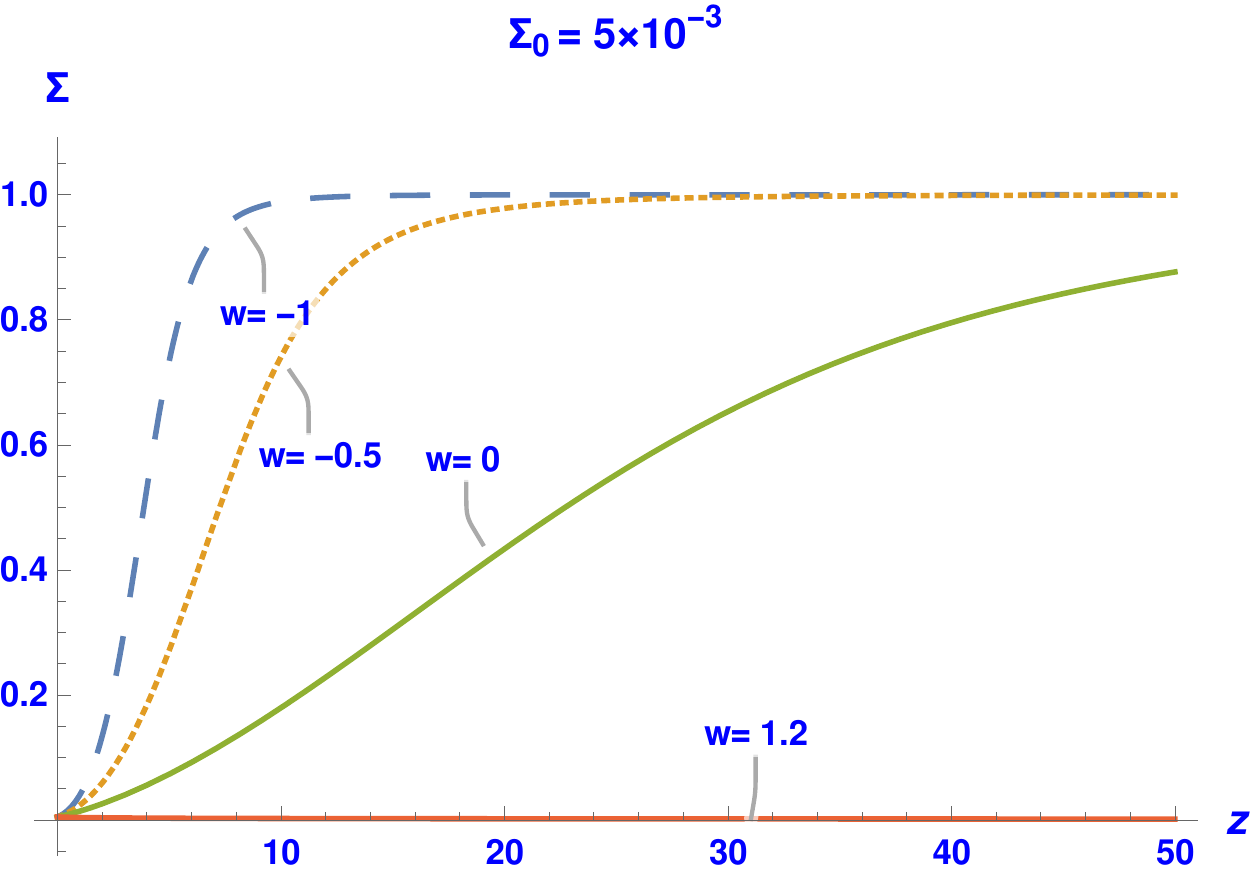}\label{Fig:perfect_fluid_plot_exemple_1}  &
    	    \includegraphics[width=0.45\textwidth,height=0.4\textwidth]{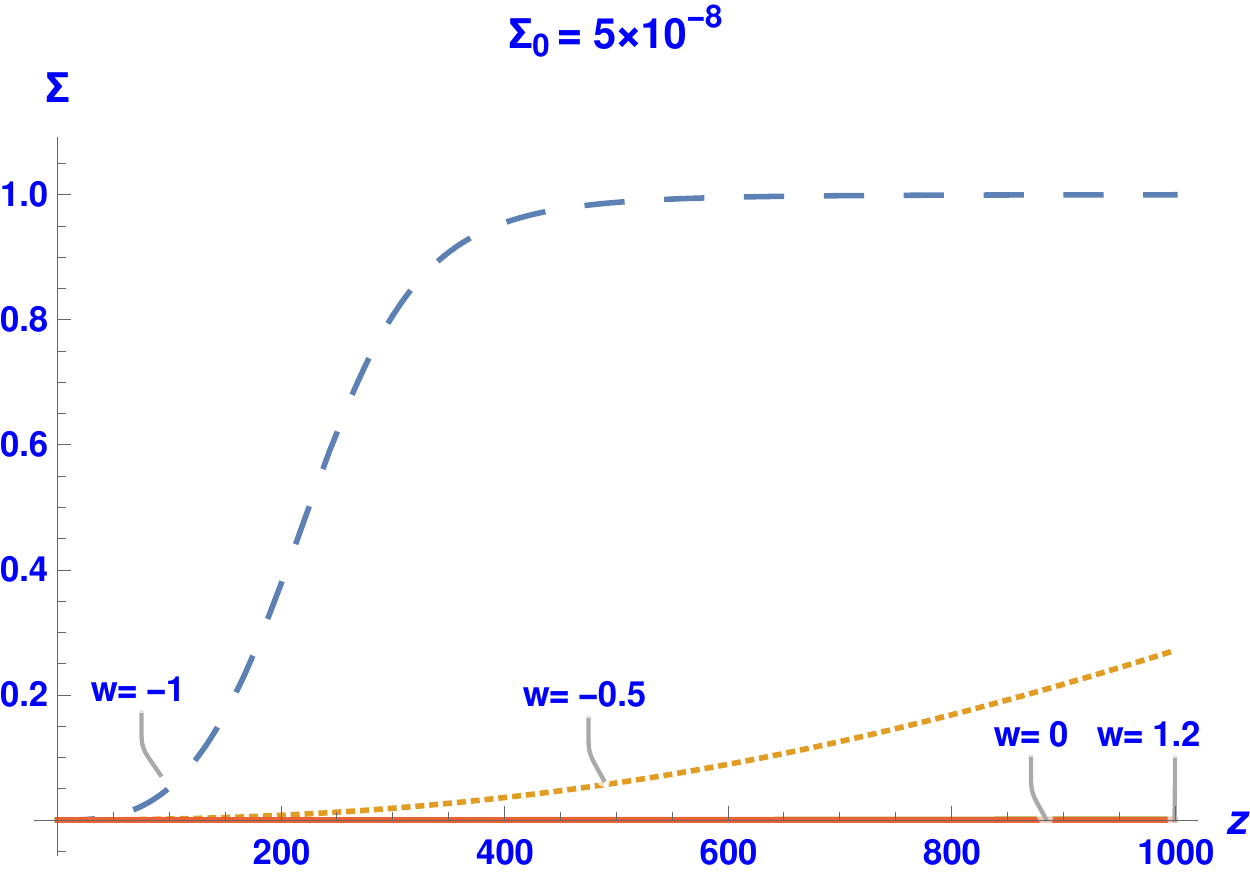}\label{Fig:perfect_fluid_plot_exemple_1}  \\
    \end{tabular}
\end{center}
The complete nonlinear analysis of the anisotropy magnitude. As $\deviation_0$ is close to the upper limits imposed by the SNe observations, the maximum anisotropy $\deviation=1$ is reached at the "small" redshifts $\isoredshift \approx 10$, for an universe dominated by the dark energy, and $\isoredshift \approx 50$ for dust. This pattern changes significantly with the order of magnitude of $\deviation_0$. The standard $\Lambda$CDM model requires $\deviation_0 \lesssim 10^{-23}$ \cite{Tedesco2019}.     
\end{table}
As it had already been noted before \cite{Tedesco2019}, if $\deviation_0 \approx 10^{-3}$ we would have to cope with the maximal anisotropy $\deviation \approx 1$ at the small redshift $\isoredshift \approx 10$, which is in complete disagreement with the standard $\Lambda$CDM model. On that grounds, we should have $\deviation_0 \lesssim 10^{-23}$. The fact is that the best estimates coming from the catalogs of type Ia supernovae at redshift $\isoredshift \lesssim 2.3$ cannot diminish the maximum anisotropy $\deviation_{Max}^{SN}$ from being somewhere in between $10^{-3}$ and $10^{-2}$, approximately 
\cite{Tedesco2019, IstLin, Campanelli, Kalus, Schucker, Wang, Chang, IstSoltis}. 
Prospects of the usage of fast radio bursts to improve this value have appeared recently \cite{Wei}. A more stringent limit is imposed if we consider the information coming from the anisotropies of the CMB, as $\deviation^{CMB}_{Max} \approx 10^{-11}$ \cite{IstSaadeh}, or even $\deviation^{CMB+BAO}_{Max} \approx 10^{-15}$ \cite{Tedesco2019}, if we use the Baryonic Acoustic Oscillations (BAO) data. They are attributed to effects at large redshift $\isoredshift \approx 10^{3}$. Clearly, these results and the success achieved by the standard model induce the strong belief that the "true" value of $\deviation_{Max}^{SN}$ should follow these estimates.

As we analyse $\factorDelta$, which is given by
\begin{equation}
\factorDelta(\isoredshift,\deviation_0) 
=  \left(\sqrt{\frac{1+ \left[(1+\isoredshift)^{3\,(1-\state_0)}-1\right]\, \deviation_0^2}{1-\deviation_0^2}\,\,} \, - \, 1 \right)
\, (1+\isoredshift)^{\frac{3}{2}\,(1+\state_0)} \, ,
\end{equation}
we note that it attains high values at redshifts of order unity, even for $\deviation_0\approx 10^{-3}$. This behavior is better seen as we define $\isoredshift_A$, the redshift for which $\factorDelta=0.01$. It roughly represents the threshold for the exponential awakening of $\factorDelta$ in the case $\state_0 \le 1$. In the dark energy sector ($w_0=-1$), $\isoredshift_A \approx 1.4$ for $\deviation_0^{DE}=0.01$, $\isoredshift_A \approx 2$ for $\deviation_0^{DE}=0.005$ and $\isoredshift_A \approx 10$ for $\deviation_0^{DE}=10^{-4}$.  If we take $\deviation_0^{Dust}\approx 2\times 10^{-8}$, for dust ($w_0=0$) , or $\deviation_0^{DE}\approx 10^{-10}$, for dark energy ($w_0=-1$), the exponential awakening would happen at $\isoredshift_A \approx \isoredshift_L \approx 1100$, where $\isoredshift_L$ is the redshift of the moment that the CMB decoupled from the matter. Hence, in the perspective of the standard $\Lambda$CDM model, $\isoredshift_A$ should be pushed forward way beyond $\isoredshift_L$.   
\begin{table}
\begin{center}
	\label{Tab:PerfectFluidBaro1}
    \begin{tabular}{cc}
    	    \includegraphics[width=0.5\textwidth,height=0.4\textwidth]{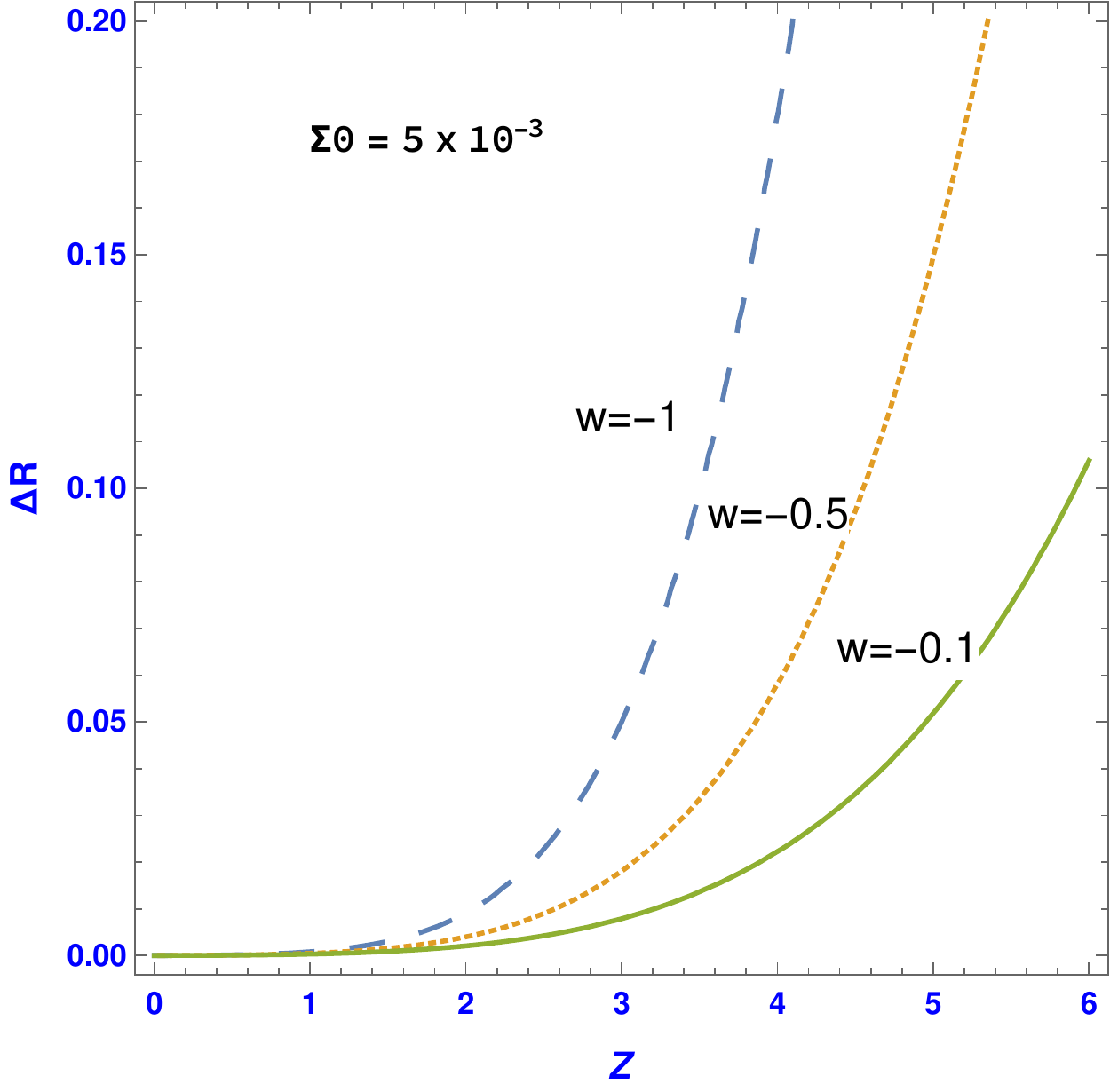}\label{Fig:perfect_fluid_plot_exemple_1}  &
    	    \includegraphics[width=0.5\textwidth,height=0.4\textwidth]{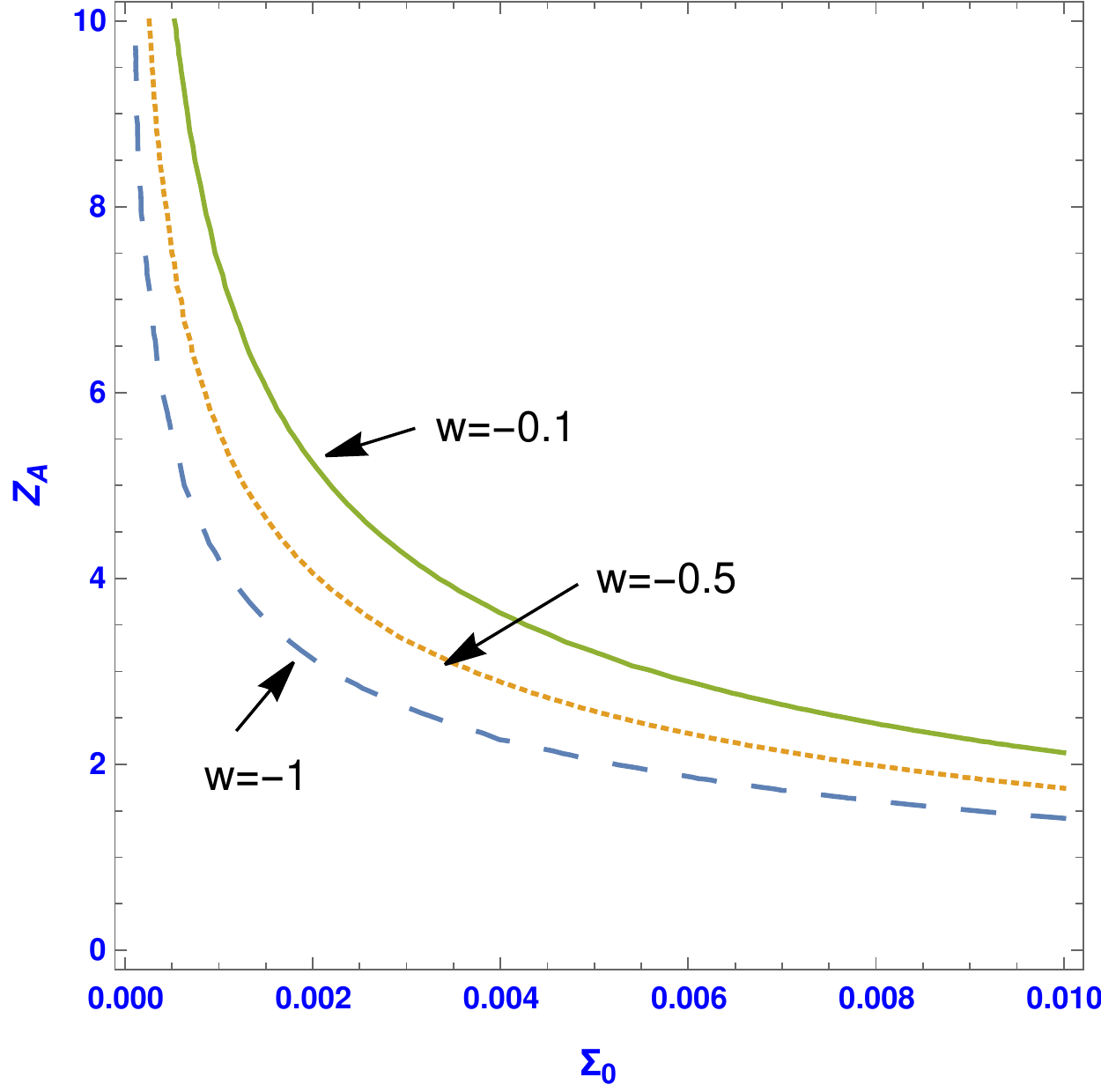}\label{Fig:perfect_fluid_plot_exemple_1} \\ 
	\end{tabular}
\end{center}
The figure in the left shows the threshold $\isoredshift_A$ of the exponential behavior of $\factorDelta$ if the anisotropy magnitude is close to the upper limits imposed by the SNe observations ($\deviation_0 \approx 5 \times 10^{-3}$). On the right, the redshift $\isoredshift_A$ is plotted in the region $0 < \deviation_0 < 10^{-2}$. Note that it gets arbitrarily large as $\deviation_0$ approaches zero.  
\end{table}

Back to the luminous distance, its distribution with the direction of observation, $\hat{n}$, involves some cumbersome analysis, so that it is hopeless to expect that some simple and general formula could arise from it \cite{Saunders,Fleury}. On the other hand, if we look along the principal directions only, things get simpler and are enough to gives us the global picture of the anisotropy's behaviour. In fact, the wave initially set to propagate parallel to a principal axis will keep its path straight \cite{Zwiebach}, that is, the geodesics along those directions are straight lines \cite{Schucker}. Hence, the respective cosmographic parameters are obtained just like in the standard model \cite{Visser}. As we define the isotropic deceleration by $q=-1-\dot{\hubble}/\hubble^2$ and use $\dot{\hubble}= -\,3\, \hubble^2 \, \left( \energy+\pressure + (\energy-\pressure)\deviation^2\, \right)/(2\energy)$, que can write $\dot{\deviation}=-\, \hubble\,\deviation\,  (2-q)$. Hence, noting that $\dot{\alpha}=0$, we obtain the expression for the deceleration along the $k$-principal direction:
\begin{equation}\label{Eq:DecelerationCalculated}
q_k =-\left(1+\frac{\dot{\hubble}_k}{\hubble_k^2}\right)_{t=t_0}= \frac{q_0 + 2\,\deviation_0\, \sin\alpha_k - 4\,\deviation_0^2\, \sin^2\alpha_k}{(1+2\,\deviation_0\, \sin\alpha_k)^2} \, ,
\end{equation}
The relation among $q_0$ and the mean value $\langle q \rangle=(q_1+q_2+q_3)/3$ is 
\begin{equation}\label{Eq:DecelerationIsotropic3}
q_0=
\langle q \rangle + 
\frac{4\,\deviation_0}{3}\, \sum_{k=1}^3 \, q_k\, \sin \alpha_k 
+ 2\,\deviation_0^2\, \left( 
1+ \frac{2}{3}\, \sum_{k=1}^3 \, q_k\, \sin^2 \alpha_k 
\right) \, .
\end{equation}
The jerk comes from $j_k = \left(1+2\, q_k \, \right)\, q_k - \dot{q}_k(t_0)/\hubble_k(t_0)$, 
\begin{equation}\label{Eq:JerkCalculated}
j_k  
=\frac{j_0+6\, \deviation_0\,\sin\alpha_k-24\, \deviation_0^2\,\sin^2\alpha_k+8\,\deviation_0^3\,\sin^3\alpha_k}{(1+2\,\deviation_0\,\sin\alpha_k)^3} \, ,
\end{equation}
with $j_0= \left(1+2\, q_0 \, \right)\, q_0 - \dot{q}(t_0)/\hubble_0$ its isotropic counterpart. Its relation with the mean value $\langle j \rangle=(j_1+j_2+j_3)/3$ is
\begin{equation}\label{Eq:JerkIsotropic3}
j_0 =  \langle j \rangle +  c_1\, \deviation_0 + \left(c_2 + 12 \right) \, \deviation_0^2 + \left(c_3  + 6\, \sin(3\,\alpha)\right)\, \frac{\deviation_0^3}{3} 
\, ,
\end{equation}
where $c_\ell = 2^\ell\,\sum_{k=1}^3\,j_k\,\sin^\ell\alpha_k$. These formulas imply the observational constraints for the anisotropy in the deceleration, 
\begin{equation}\label{Eq:ConstraintQk}
q_k = \left(\frac{\hubble_0}{\hubble_k}\right)^2\,q_0 - \left(1-\frac{\hubble_0}{\hubble_k}\right)\left(1-2\,\frac{\hubble_0}{\hubble_k}\right)\, . 
\end{equation}
and in the jerk,
\begin{equation}\label{Eq:ConstraintJk}
j_k = \left(\frac{\hubble_0}{\hubble_k}\right)^3\,j_0 + \left(1-\frac{\hubble_0}{\hubble_k}\right)\left(1-8\,\frac{\hubble_0}{\hubble_k}+10 \, \left(\frac{\hubble_0}{\hubble_k}\right)^2\right)\, . 
\end{equation}
Note that the formulas above are complete, no approximation method have been used, and they reduce to a tautology if the universe is isotropic ($\deviation_0=0$). We can resume them in the following theorem:
\begin{thm}\label{Thm:ConstraintQJ}
Under the hypothesis of the theorem \ref{Thm:ConstraintHubble}, the deceleration and the jerk along the principal directions satisfy the algebraic constraints (\ref{Eq:ConstraintQk}) and (\ref{Eq:ConstraintJk}). In the linear regime, they are expressed as $q_0\approx\langle q \rangle$, $j_0\approx\langle j \rangle$, 
\begin{equation}\label{Eq:ConstraintQikLinear}
\delta q_k \approx  \left(1-2\langle q \rangle\right)\, 
\frac{\delta\hubble_k}{\hubble_0}  
\quad \textrm{and} \quad 
\delta j_k \approx 3\, \left(
1 -  \langle j \rangle   
\right)\, 
\frac{\delta\hubble_k}{\hubble_0}  
 \, ,
\end{equation}
where $\delta \hubble_k \ll \hubble_0$, with $\delta \hubble_k = \hubble_k-\hubble_0$, $\delta q_k = q_k-\langle q \rangle$ and $\delta j_k = j_k-\langle j \rangle$.         
\end{thm}
It is quite interesting that the constraints (\ref{Eq:ConstraintQikLinear}) for the cosmological anisotropy define very simple patterns for $\delta q_k/\delta \hubble_k$ and $\delta j_k/\delta \hubble_k$, such that they do not depend on the direction. Hence, any claim of the existence of the anisotropy in the deceleration, as for instance $\delta q_1 \approx 10^{-2}$ \cite{AnstCai}, should be accompanied by the observation of the first constraint in (\ref{Eq:ConstraintQikLinear}) along all the three principal directions.

Despite the prospects created by the very small value of $\deviation_0$ in the CMB, the direct search for the upper bounds for the cosmological anisotropy in the late time sky, or even its unexpected detection, is an important and independent check for the standard $\Lambda$CDM model. We recall that, at the moment of writing, there is a still unresolved tension between the measurement of the Hubble constant $\hubble_0^{SNe}$ from the Hubble space telescope and that $\hubble_0^{CMB}$ from the Planck data \cite{Riess}, calling our attention for the fact that the observations of the CMB and the SNe might not agree even at $\isoredshift < 0.15$. Hence, another test for the compatibility of the standard model with these two independent data sets is important for the theory. At this point, the theorems \ref{Thm:ConstraintHubble} and \ref{Thm:ConstraintQJ} come as useful aides, for they will be of valuable importance in the interpretation of the nature of the anisotropy that might appear in the  supernovae surveys \cite{Schucker, AnstCai, AnstColin}.


\vspace{1cm}

\begin{thebibliography}{00}  
%
\bibitem{ellis_mac_marteens}
Ellis G F R , Maartens R and MacCallum M A H 2012 {\em Relativistic Cosmology} (Cambridge University Press)
%
\bibitem{wainwright}
Wainwright J and Ellis G F R 2009 {\em Dynamical Systems in Cosmology} (Cambridge University Press) 
%
\bibitem{Tedesco}
Tedesco L 2018 Ellipsoidal expansion of the universe, cosmic shear, acceleration and jerk parameter  {\em Eur. Phys. J. Plus} {\bf 133}  188 [gr-qc/1804.11203v1]
%
\bibitem{BGK}
Bittencourt E, Gomes L G and Klippert R 2017 Bianchi-I cosmology from causal thermodynamics {\it Class. Quantum Grav.} {\bf  34} 045010
%
\bibitem{Weinberg}
Weinberg S 2008 {\it Cosmology}, (Oxford University Press)
%
\bibitem{BGS}
Bittencourt E, Gomes L G, dos Santos G B 2021 Intrinsically symmetric cosmological model in the presence of dissipative fluids {\it Int. J. Mod. Phys. D}  {\bf ...}
%
\bibitem{Linder}
Denissenya M, Linder E V and Shafieloo A 2018 Cosmic curvature tested directly from observations {\it JCAP03} 041
%
\bibitem{Ellis_MacCallum}
Ellis G F H , MacCallum M A H 1969 A class of homogeneous cosmological models
{\em Commun. math. phys.} {\bf 12} 108-141 
%
\bibitem{Tsoubelis}
Tsoubelis D 1979 Bianchi VI0, VII0 cosmological models with spin and torsion  {\em PRD} {\bf 20} 3004-3008  
%
\bibitem{Tedesco2019}
Akarzu O, Kumar S, Sharma S and Tedesco L 2019 Constraints on Bianchi type I spacetime extension of the standard $\Lambda$CDM model {\em PRD} {\bf 100} 023532.
%
\bibitem{IstLin}
Lin H, Wang S, Chang Z and Li X 2016 Testing the isotropy of the Universe by using the JLA
compilation of type-Ia supernovae {\it Mon. Not. R. Astron. Soc.} {\bf 456} Issue 2, 1881–1885 [astro-ph/1504.03428v2]
%
\bibitem{Campanelli}
Campanelli L, Cea P, Fogli G L and  Marrone A 2011 Testing the isotropy of the universe with type Ia supernovae {\em PRD} {\bf 83}  103503.
%
\bibitem{Kalus}
Kalus B , Schwarz D J, Seikel M and Wiegand A 2013 Constraints on anisotropic cosmic expansion from supernovae {\em  A$\&$A} {\bf 553} A56
%
\bibitem{Schucker}
Schucker T , Tilquin A and Valent G 2014 Bianchi I meets the Hubble diagram {\em Mon.Not.Roy.Astron.Soc.} {\bf 444} 2820-2836.
%
\bibitem{Wang}
Wang Y Y and Wang F Y 2018 Testing the isotropy of the Universe with type Ia supernovae in a model-independent way {\em  Mon.Not.Roy.Astron.Soc.} {\bf 474} 3, 3516-3522
%
\bibitem{Chang}
Zhao D, Zhou Y and Chang Z 2019 Anisotropy of the Universe via the Pantheon supernovae sample revisited {\em Mon.Not.Roy.Astron.Soc.} {\bf 486} 4, 5679–5689
%
\bibitem{IstSoltis}
Soltis J, Farahi A, Huterer D, Liberato II C M 2019 Percent-Level Test of Isotropic Expansion Using Type Ia Supernovae  {\em Phys. Rev. Lett.} {\bf 122} 091301 [astro-ph/1902.07189 ]
%
\bibitem{Wei}
Qiang D-C,  Deng H-K and Wei H 2020 Cosmic anisotropy and fast radio bursts {\em Class. Quantum Grav.} {\bf 37}
%
\bibitem{IstSaadeh}
Saadeh D, Feeney S M, Pontzen A, Peiris H V and McEwen J D 2016 How Isotropic is the Universe? {\em Phys. Rev. Lett.} {\bf 117} 131302 [astro-ph/1605.07178]
%
\bibitem{Saunders}
Saunders P T 1969 Observations in some simple cosmological models with shear {\em  Mon. Not. Roy. Astron. Soc.} {\bf 142} 213
%
\bibitem{Fleury}
Fleury P, Pitrou C, Uzan J P 2015 Light propagation in a homogeneous and anisotropic universe {\em PRD} {\bf 91} 043511
%
\bibitem{Zwiebach}
Sagnotti A and Zwiebach B 1981 Electromagnetic waves in a Bianchi type-I universe {\em Phys. Rev. D} {\bf 24} 305 
%
\bibitem{Visser}
Visser M 2004 Jerk, snap and the cosmological equation of state {\em Class. Quantum Grav.} {\bf 21}   2603 [gr-qc/0309109v4]
%
\bibitem{Riess}
Riess A G et al., {\em A $2.4\%$ determination of the local value of the Hubble constant}, 
The Astrophysical Journal, {\bf 826} (2016) 56
doi:10.3847/0004-637X/826/1/56
%
\bibitem{AnstCai}
Cai R G and Tuo Z L 2012 Direction dependence of the deceleration parameter {\em JCAP02} 004 [astro-ph/1109.0941]
%
\bibitem{AnstColin}
Colin J, Mohayaee R, Rameez M and  Sarkar S 2019 Evidence for anisotropy of cosmic acceleration {\em Astron. $\&$ Astrophysics} {\bf 631} L13 [astro-ph/1808.04597v3]


%
\end{thebibliography}
\end{document}